\documentstyle[aps,prl,multicol]{revtex}
\begin{document}
\draft
\title{Self-diffusion in  granular gases}
\author{Nikolai V. Brilliantov$^{1,2}$ and  Thorsten P\"oschel$^{2}$ }

\address{$^1$Moscow State University, Physics Department, Moscow
  119899, Russia}
\address{$^2$Humboldt-Universit\"at, Institut f\"ur Physik,
  Invalidenstr. 110, D-10115 Berlin, Germany}
   
\date{\today}
\maketitle
\begin{abstract}
  The coefficient of self-diffusion for a homogeneously cooling
  granular gas changes significantly if the impact-velocity dependence
  of the restitution coefficient $\epsilon$ is taken into account. For
  the case of a constant $\epsilon$ the particles spread
  logarithmically slow with time, whereas the velocity dependent
  coefficient yields a power law time-dependence. The impact of the
  difference in these time dependences on the properties of a freely
  cooling granular gas is discussed.
\end{abstract}


\begin{multicols}{2}

\section{Introduction}

The behavior of freely evolving granular gas has been intensively
discussed recently, in particular the process of cluster formation due
to inelastic collisions was of wide interest,
e.g.~\cite{GZ93,McNamara,Cluster}. On the basis of a continuum description the
effect of clustering in a force-free granular gas has been explained
as an instability of the hydrodynamic equations~\cite{GZ93,McNamara,Cluster}.
For a deeper understanding of clustering phenomena it may be worth to
consider the processes in a granular gas which {\em precede}
clustering. To this end we investigate the effect of self-diffusion of
particles in the regime of homogeneous cooling.

The collisions of particles in a granular gas can be described by the
coefficient of restitution $\epsilon$ which relates the normal
component of the relative velocity 
$\vec{v}_{\mbox{\footnotesize\em  ij}}=\vec{v}_{i}-\vec{v}_{j}$ before 
a collision to that after the
collision $\vec{v}_{\mbox{\footnotesize\em ij}}^{\,\prime}$ as
$|\vec{v}_{\mbox{\footnotesize\em ij}}^{\,\prime} \vec{e}| = \epsilon
|\vec{v}_{\mbox{\footnotesize\em ij}} \vec{e}|$.  The unit vector
$\vec{e}=\vec{r}_{\mbox{\footnotesize\em
    ij}}/|\vec{r}_{\mbox{\footnotesize\em ij}}|$ gives the direction
of $\vec{r}_{\mbox{\footnotesize\em ij}}=\vec{r}_i-\vec{r}_j$ at the
instant of the collision.

Up to our knowledge, all analytical calculations for the force-free
case reported so far refer to systems of particles colliding with a constant
restitution coefficient $\epsilon$. Experiments as well as theoretical
studies show, however, that $\epsilon$ depends on the normal component 
of the impact velocity
$|\vec{v}_{\mbox{\footnotesize\em ij}}\vec{e}|$~\cite{CollExp}.

The problem of the restitution coefficient's dependence on the impact
velocity has been addressed in~\cite{BSHP}, where the generalization
of the Hertz contact problem was developed for the collision of
viscoelastic particles.  From this generalized Hertz equation one
obtains the velocity-dependent restitution coefficient~\cite{TomThor}
as an expansion:
\begin{eqnarray}
\epsilon&=&1- C_1\!\left(\frac{3A}{2} \right) \alpha^{2/5}\left|\vec{e}\,
\vec{v}_{\mbox{\footnotesize\em ij}}\right|^{1/5} \nonumber \\
&+& C_2 \left(\frac{3A}{2} \right)^2  \alpha^{4/5}\left|\vec{e}\,
\vec{v}_{\mbox{\footnotesize\em ij}}\right|^{2/5}  \mp  \cdots
\label{epsilon}
\end{eqnarray}
with 
\begin{equation}
\alpha= 
\frac{ 2~ Y\sqrt{R^{\,\mbox{\footnotesize\em eff}}}}{
    3~ m^{\mbox{\footnotesize\em eff}}\left( 1-\nu ^2\right) }
\end{equation}
where $Y$ is the Young modulus, $\nu$ is the Poisson ratio, and $A$
depends on dissipative parameters of the particle material 
(for details see~\cite{BSHP}). The effective mass and radius is defined as
\begin{eqnarray}
R^{\,\mbox{\footnotesize\em eff}}&=&R_1R_2/(R_1+R_2)\\
m^{\mbox{\footnotesize\em eff}}&=&m_1m_2/(m_1+m_2)
\end{eqnarray}
with $R_{1/2}$ and $m_{1/2}$ being the radii and masses of the
colliding particles. The constants $C_1=1.15344$ and $ C_2=0.79826 $
were obtained analytically \cite{TomThor}, confirmed then by numerical
simulations and may be also written in a closed form as
\cite{Rozaetal}

\begin{eqnarray}
\label{C1}
C_1&=&\frac{ \Gamma(3/5)\sqrt{\pi}}{2^{1/5}5^{2/5} \Gamma(21/10)}\\
C_2&=&\frac35C_1^2   
\label{C2}
\end{eqnarray}
Eq.(\ref{epsilon}) refers to the case of pure viscoelastic
interaction, i.e. when the relative velocities $|\vec{v}_{ij}\vec{e}|$
are not too large (to avoid plastic deformation of the particles) and are 
not too small (to allow to neglect surface effects such as roughness,
adhesion and van der Waals interactions). This implies that the initial 
temperature of the granular gas is not too large and the final temperature
is not too small. The range of validity of Eq.(\ref{epsilon}) depends on 
material and surface properties. In spite of such restrictions some 
important systems in nature do exist (e.g. planetary rings), where these 
conditions are satisfied \cite{remarkexp}. Here we assume that the 
granular gas conditions allows for the application of Eq.(\ref{epsilon}).

For equilibrium 3D-system the time-dependence of the mean-square
displacement reads
\begin{equation}
\left\langle \left( \Delta r(t) \right)^2  \right\rangle_{\rm eq} 
= 6\,D\,t \,.
\label{Dgen}
\end{equation}
where $\left< \cdots \right>_{\rm eq}$ denotes the {\em equilibrium}
ensemble averaging. 
To calculate the mean-square  displacement, one writes 
\begin{equation}
\left\langle \left( \Delta r(t) \right)^2  \right\rangle_{\rm eq} =
\left< \int_0^{t}\vec{v}(t^{\prime})dt^{\prime}
\int_0^t  \vec{v}(t^{\prime \prime}) dt^{\prime \prime}  \right>_{\rm eq} \, 
\label{delR}
\end{equation}
and encounters then with the velocity autocorrelation function,
$\left< \vec{v}(t^{\prime}) \vec{v}(t^{\prime \prime}) \right>_{\rm eq} $.  
For systems at equilibrium this depends only on the time
difference, $|t^{\prime}-t^{\prime \prime}|$ and decays with a
characteristic time $\tau_v$. Therefore, at time $t \gg \tau_v$ one
arrives at the self-diffusion coefficient
\begin{equation}
D=\frac13 \, \int_0^{\infty}\,
\left\langle \vec{v}(0) \vec{v} (t)\,\right\rangle_{\rm eq} dt \,.
\label{Dvel}
\end{equation}

For {\em nonequilibrium} systems such as granular materials the
concept of the self-diffusion coefficient may be also applied, but with
some care and with necessary generalization. Obviously, this refers
only to ``liquid'' or gaseous phases of the
material~\cite{EsipovPoeschel:97} where the particles have a
noticeable mobility. In the following we also restrict ourselves to
homogeneous cooling and consider the stages of evolution preceding the
cluster formation~\cite{GZ93,Cluster}, i.e.  we assume that the
granular material is (at least locally) homogeneous and isotropic.
Hence, one can define the temperature $T(t)$, which decreases with
time due to the loss of energy according to inelastic collisions. For the
impact-velocity dependent restitution coefficient (\ref{epsilon}) one
has \cite{TomThor}:

\begin{equation}
T =  T_0/(1+t/\tau_0)^{5/3}
\label{temp}
\end{equation} 
where $T_0$ is the initial temperature and $\tau_0$ is the
characteristic time of the cooling process, which may be estimated as
\begin{equation}
\tau_0^{-1} \sim \sigma^2 n A \alpha^{2/5} T_0^{3/5}  
\end{equation}
with $\sigma=2R$ and $n$ being the particle diameter and the particle
number density, respectively.  The mean-collision time
\begin{equation}
\tau_c^{-1}(t) \equiv 4 \pi^{1/2} g_2(\sigma)
n \sigma^2 T^{1/2}
\end{equation} 
depends on time via the time-dependent temperature. Here $g_2(\sigma)
$ is the contact value of the two-particle radial distribution
function and the particles are of unit mass. Thus, the ratio of the
two characteristic times reads

\begin{equation}
\tau_c(t)/\tau_0 \sim \delta^{11/6}\left[t/\tau_c(0)\right]^{5/6}
\label{tauratio}
\end{equation} 
where $\tau_c(0)^{-1}$ is the collisional frequency at initial time,
and $\delta=A\alpha^{2/5}T_0^{1/10}$ is supposed to be small.
Clearly, one can employ the concept of temperature if
$\tau_c(t)/\tau_0 \ll 1$. Thus, Eq.~(\ref{tauratio}) gives the estimate
$t \ll \tau_c(0) \delta^{-11/5} $ for the upper time limit for which the
use of the local temperature is justified.

An important property of homogeneously cooling granular gas is that
the velocity distribution is close to Maxwellian. Moreover, it
persists with time, changing in accordance with the decreasing
temperature~\cite{EsipovPoeschel:97}. The small value of the fourth
cumulant of the velocity distribution function for any value of the
restitution coefficient reported in~\cite{NoijeErnst:97} also supports
the Maxwellian distribution~\cite{NoijeErnst:97,NoijeErnst:98}.
Therefore, we assume that the Maxwellian distribution and the
molecular chaos hypothesis may be used with a good degree of accuracy.

The evolution on the hydrodynamic time-scale may be described using
the kinetic coefficients calculated on the short-time scale $t\sim
\tau_c$. For granular gases these transport coefficient will be now
time-dependent.  We calculate the self-diffusion coefficient within
the uncorrelated binary collisions approximation and assume that the
inequality $\tau_c(t) \ll \tau_0$ always holds true (see the above
discussion on this condition).  Thus, on the time-scale $t\sim \tau_c$
the temperature may be considered as a constant. For $t \gg \tau_c$,
however, the self-diffusion coefficient occurs to be time-dependent
and the generalization of Eq.~(\ref{Dgen}) reads
\begin{equation}
\left\langle \left( \Delta r(t) \right)^2  \right\rangle =
6\, \int^t D(t^{\prime}) dt^{\prime}
\label{Dgengen}
\end{equation}
where $\langle \cdots \rangle$ denotes averaging over the {\em
  nonequilibrium} ensemble, which evolution is described by a
time-dependent $N$-particle distribution function $\rho(t)$.  Here we
restrict ourselves to times when the hydrodynamic contribution to the
self-diffusion coefficient is not large~\cite{remhydcon}, so that
(\ref{Dgengen}) with $D(t)$ calculated on the time scale $t\sim
\tau_c$ gives an accurate description of the mean-square displacement.

The aim of the present study is to analyze how the velocity dependence of 
the restitution coefficient influences the diffusion in a gas of identical 
particles. The paper is organized as follows: In next Sec.II we obtain the 
time-dependence of the diffusion coefficient and the temperature of the granular 
gas in homogeneous cooling state. We also show that the mean-square displacement
depends on time quite differently for the case of the constant restitution 
coefficient and for that determined by the impact velocity. In Sec.III we 
discuss our results for the mean-square displacement in a context of its 
possible impact on clustering.

\section{Time dependence of the diffusion coefficient and of temperature}

To describe the dynamics of the granular material we use the formalism
of the pseudo-Liouville operator ${\cal L}$~\cite{pseudo}

\begin{equation}
i{\cal L}=\sum_j \vec{v}_j \cdot \frac{\partial}{\partial \vec{r}_j}
+\sum_{i<j}\, \hat{T}_{\mbox{\footnotesize\em ij}}\,.
\label{L}
\end{equation}
The first sum in (\ref{L}) refers to the free streaming of the
particles (the ideal part) while the second sum refers to the particle
interactions which are described by the binary collision
operators~\cite{Chandler}
\begin{equation}
\hat{T}_{\mbox{\footnotesize\em ij}}\!=\sigma^{2}\!\! \int\!\! d^2\vec{e}\, 
\Theta \left(- \vec{v}_{\mbox{\footnotesize\em ij}} \cdot \vec{e}\, \right)\!
|\vec{v}_{\mbox{\footnotesize\em ij}} \cdot \vec{e}\, | 
\delta\! \left( \vec{r}_{\mbox{\footnotesize\em ij}}- \sigma \vec{e} 
\right)\!\!
\left(\hat{b}_{\mbox{\footnotesize\em ij}}^{\vec{e}}-1 \right)  
\label{Tij}
\end{equation}
where $\Theta(x)$ is the Heaviside function.  The operator
$\hat{b}_{\mbox{\footnotesize\em ij}}^{{\vec{e}}}$ is defined as

\begin{equation}
\hat{b}_{\mbox{\footnotesize\em ij}}^{\vec{e}} f \left (\vec{r}_{i},
  \vec{r}_{j}, \vec{v}_{i},\vec{v}_{j} \cdots \right)=f \left
  (\vec{r}_{i}, \vec{r}_{j},
  \vec{v}^{\,\prime}_{i},\vec{v}^{\,\prime}_{j} \cdots \right) \, , 
\end{equation}
where $f$ is some function of dynamical variables.  The
after-collision velocities of the colliding particles,
$\vec{v}^{\,\prime}_{i}$ and $\vec{v}^{\,\prime}_{j}$ are related to
their pre-collisional values $\vec{v}_{i}$, $\vec{v}_{j}$ via

\begin{equation}
\vec{v}_{i,j}^{\,\prime} = \vec{v}_{i,j} \mp
\frac12 \,(1+\epsilon)\left(\vec{v}_{\mbox{\footnotesize\em ij}} \cdot
  \vec{e}\,\right) \vec{e}\, .
\end{equation}

The pseudo-Liouville operator gives the time derivative of any
dynamical variable $B$ (e.g. \cite{resibua}):
\begin{equation}
\frac{d}{dt} B\left( \left\{ \vec{r}_i, \vec{v}_i  \right\}, t \right)= 
i{\cal L}\, B\left( \left\{     \vec{r}_i, \vec{v}_i \right\}, t \right)
\label{derA}
\end{equation}
and, therefore, the  time evolution of $B$
reads ($t>t^{\prime}$)
\begin{equation}
B\left( \{ \vec{r}_i, \vec{v}_i  \}, t \right)=
e^{i{\cal L} (t-t^{\,\prime} )} 
B\left( \{     \vec{r}_i, \vec{v}_i \}, t^{\,\prime} \right)\,.
\label{evolA}
\end{equation}
With (\ref{evolA}) 
the time-correlation function reads
\begin{equation}
\left< B(t^{\prime})B(t) \right >=
\int d\Gamma \rho(t^{\prime}) B(t^{\prime}) e^{i{\cal L} (t-t^{\prime})}
B(t^{\prime})\,,
\label{evolAA}
\end{equation}
where $\int d\Gamma$ denotes integration over all degrees of freedom
and $\rho(t^{\prime})$ depends on temperature $T$, density $n$, etc.,
which change on a time-scale $t \gg \tau_c$. In accordance with the
molecular chaos assumption at $ t \sim \tau_c$ the sequence of the
successive collisions occurs without correlations. If the variable $B$
does not depend on the positions of the particles, its
time-correlation function reads~\cite{Chandler1}
\begin{equation}
\left\langle B(t^{\prime})B(t) \right \rangle =
\left< B^2 \right>_{t^{\prime}} e^{-\left|t-t^\prime\right|/\tau_B(t^\prime)}
~~~~\left(t > t^{\prime}\right) \,.
\label{AAexp}
\end{equation}
where $\langle \cdots \rangle_{t^\prime}$ denotes the averaging with
the distribution function taken at time $t^{\prime}$.  The relaxation
time $\tau_B$ is inverse to the initial slope of the autocorrelation
function~\cite{Chandler1}.  It may be found from the time derivative
of $\left\langle B(t^{\prime} )B(t) \right \rangle$ taken at
$t=t^{\prime}$. Eqs.~(\ref{evolAA}) and (\ref{AAexp}) then yield
\begin{equation}
-\tau_B^{-1}(t^{\prime})=\int\!\!d\Gamma \rho(t^{\prime}) B {i{\cal L} }B /
\left\langle B^2 \right \rangle_{t^{\prime}}= 
\frac{\left\langle B i{\cal L}  B \right 
\rangle_{t^{\prime}}}{\left\langle B^2 \right \rangle_{t^{\prime}}}\,.
\label{ALA}
\end{equation}
The relaxation time $\tau_B^{-1}(t^{\prime})$ depending on time via
the distribution function $\rho(t^{\prime})$, changes on the
time-scale $t \gg \tau_c$.

Let $B(t)$ be the velocity of some particle, say $\vec{v}_1(t)$.  Then
with $3T(t)=\left\langle v^2 \right \rangle_t$
Eqs.~(\ref{AAexp},\ref{ALA}) (with (\ref{L},\ref{Tij})) read
\begin{equation}
\left\langle \vec{v}_1 (t^{\prime})\cdot  \vec{v}_1(t)\right\rangle =
3T(t^{\prime}) e^{-|t-t^{\prime}|/\tau_v(t^{\prime})}
\label{vvexp}
\end{equation} 
\begin{equation}
-\tau_v^{-1}(t^\prime)=
(N-1) \frac{\left< \vec{v}_1 \cdot \hat{T}_{12} \vec{v}_1\right>_{t^{\prime}}}{\left< \vec{v}_1  \cdot \vec{v}_1 \right>_{t^{\prime}}}\,.
\label{vTv}
\end{equation} 
To obtain (\ref{vTv}) we take into account that ${\cal L}_0
\vec{v}_1=0$, $\hat{T}_{\mbox{\footnotesize\em ij}}\vec{v}_1=0$ if $i
\neq 1$ and the identity of the particles.  The calculation of
$\tau_v^{-1}(t^\prime)$ may be performed if we assume that the
distribution function $\rho(t^{\prime})$ is a product of the
coordinate part, which corresponds to a uniform and isotropic system,
and a velocity part which is a product of Maxwellian distribution
functions 
\begin{equation}
\phi(\vec{v}_i)=\frac{\exp[-v_i^2/2T(t^{\prime})]}{[2\pi
T(t^{\prime})]^{3/2}}\,, \quad i=1, \ldots, N\, .
\nonumber 
\end{equation} 
Integration over the coordinate part in (\ref{vTv}) yields
\begin{equation}
(N\!-\!1)\!\int\!\rho(t^{\prime})
\delta \left(\vec{r}_{\mbox{\footnotesize\em ij}}\!-\!\sigma 
\vec{e} \right) d\vec{r}_1 \cdots d\vec{r}_N\
\!=\!ng_2(\sigma) \prod_i \phi(\vec{v}_i )\,,  
\end{equation}
where we use the definition of the configurational distribution
functions \cite{resibua}, and where \cite{resibua,Chandler1}
\begin{equation}
g_2(\sigma)=\frac12 (2-\eta)/(1-\eta)^3  
\end{equation}
gives the contact value of the configurational distribution function 
and $\eta=\frac16 \pi n\sigma^3 $.
 With 
 \begin{equation}
\left< \vec{v}_1 \hat{T}_{12}
  \vec{v}_1 \right>_{t^{\prime}} =\frac12 \left< \vec{v}_{12}
  \hat{T}_{12} \vec{v}_1 \right>_{t^{\prime}}   
 \end{equation}
due to the collision rules and definition (\ref{Tij}), one finally arrives at
\begin{eqnarray}
&&\tau_v^{-1}(t^\prime) = \frac14 ng_2(\sigma) \sigma^{2} 
\int d\vec{v}_{12} \phi (\vec{v}_{12} ) \label{vTv2}\\
&&~~ \int d^2 \vec{e}  \Theta(-\vec{v}_{12} \cdot \vec{e}) 
|\vec{v}_{12} \cdot \vec{e}|\,(\vec{v}_{12} \cdot \vec{e})^2(1+\epsilon)/ 
\left< \vec{v}_1 \cdot  \vec{v}_1\right >_{t^{\prime}} \nonumber
\end{eqnarray}
where 
\begin{equation}
\phi (\vec{v}_{12} )=(4\pi T)^{-3/2} \exp(-v_{12}^2/4T)  
\end{equation}
is the
Maxwellian distribution for the relative velocity of two particles.
For $\epsilon$ not depending on $v_{12}$ Eq.~(\ref{vTv2}) yields
\begin{equation}
\tau_v^{-1}(t)=\frac{\epsilon +1}{2}\frac83 n \sigma^2 g_2(\sigma) 
\sqrt{\pi T(t)} =\frac{\epsilon +1}{2}  \tau_E^{-1}(t)\,, 
\label{tEt}
\end{equation}
where $\tau_E(t)=\frac32\,\tau_c(t)$ is the Enskog relaxation time
\cite{resibua}. For the granular gas it depends on time according to
the same time-scale as the temperature. As shown in~(\ref{tEt}) the
velocity correlation time for inelastic collisions exceeds that of
elastic ones. This follows from partial suppression of the
backscattering of particles due to inelastic losses in their normal
relative motion. As a result the trajectories of particles are more
stretched as compared with the elastic case, and therefore the velocity
correlation time is larger.

As discussed above a constant restitution coefficient is not
consistent with the nature of the inelastic collisions. Substituting
(\ref{epsilon}) into (\ref{vTv2}) one finds the velocity correlation
time for the gas of inelastically colliding spheres:
\begin{eqnarray}
\tau_v^{-1}(t)&=&\tau_E^{-1}(t) \left[1-\frac34 
\Gamma\left( \frac{21}{10}  \right) 
C_1 A \alpha^{2/5}\left(4T(t) \right)^{1/10} +\right.\nonumber \\
&+& \left. \frac{27}{40}
\Gamma\left( \frac{11}{5} \right) C_1^2 A^2 
\alpha^{4/5} \left( 4T(t) \right)^{1/5} \mp \cdots 
\right]\,,
\label{tfin}
\end{eqnarray}
where $\Gamma(x)$ is the Gamma-function, $\tau_E$ is given by
(\ref{tEt})~\cite{expansion} and we use (\ref{C2}), which 
relates the coefficients $C_1$ and $C_2$.  From (\ref{tfin}) follows that the
velocity autocorrelation function decays (as expected) on the short
time scale, since $\tau_v$ is of the order of $\tau_c$.

Using the velocity correlation function one writes
\begin{equation}
\left\langle \left( \Delta r(t) \right)^2  \right\rangle 
=2 \int_0^t dt^\prime 3T (t^\prime) \int_{t^\prime}^t 
dt^{\prime\prime} e^{-|t^{\prime\prime}-t^\prime|/\tau_v(t^\prime)}\,.
\label{Difvel}
\end{equation}
On the short-time scale $t \sim \tau_c$, $T (t^\prime)$ and
$\tau_v(t^\prime)$ may be considered as constants. Integrating in
(\ref{Difvel}) over $t^{\prime\prime}$ and equating with
(\ref{Dgengen}) yields for $t \gg \tau_c \sim \tau_v$ the
time-dependent self-diffusion coefficient
\begin{equation}
D(t)= T(t) \tau_v(t)\,.
\label{Difviatau}
\end{equation}

Using the pseudo-Liouville operator one can also describe the
time-dependence of the temperature of the granular gas with the
impact-velocity dependent restitution coefficient. From (\ref{derA})
it follows (see also \cite{HuthmannZippelius:97}):
\begin{equation}
\dot{T}(t)=\frac13\, \frac{d}{dt} \left< v^2 \right>_t=
\frac13\,  \left< i {\cal L} \,v^2 \right>_t \, .  
\end{equation}
Calculations similar to that for  $\tau_v(t^{\prime})$ yield
\begin{eqnarray}
\dot{T}&=&-b_1T^{8/5}+b_2T^{17/10} \mp  \cdots \label{Tdot1}\\
b_1&=&4\cdot 2^{1/5} \pi^{1/2} \Gamma 
\left( \frac{21}{10} \right) C_1
\sigma^2ng_2(\sigma)A \alpha^{2/5}\nonumber\\
b_2&=&\frac{33}{5} \,2^{2/5} \pi^{1/2} 
\Gamma \left( \frac{11}{5} \right)C_1^2 
\sigma^2ng_2(\sigma)A^2  \alpha^{4/5}\nonumber
\end{eqnarray}
Solving Eq.(\ref{Tdot1}) and expanding the result in terms of the
small parameter 
\begin{equation}
\delta=A\alpha^{2/5}T_0^{1/10}  
\end{equation}
one arrives at
\begin{equation}
\frac{T(t)}{T_0}\!=\!\left(\!1\!+\!\frac{t}{\tau_0}
\right)^{\!\!-\frac{5}{3}} \!\!\!\!\!+ a_1 \, \delta
\left(\!1\!+\!\frac{t}{\tau_0}\right)^{\!\!-\frac{11}{6}}\!\!\!\!\!+ a_2 
\delta^2 \, \left(\!1\!+\!\frac{t}{\tau_0}\right)^{\!\!-2}\!\!\!\!\! + 
\cdots
\label{Tres}
\end{equation}
with $a_1$ and $a_2$ being pure numbers~\cite{constants} and with 
\begin{eqnarray}
\tau_0^{-1} &=& \delta \cdot \tau_c(0)^{-1} \cdot 
\frac35 \, 2^{1/5} \,C_1
\Gamma\left( \frac{21}{10} \right)\nonumber\\
&=& 0.831928 \cdot \delta \cdot \tau_c(0)^{-1}  
\end{eqnarray}
The leading term in this
expansion corresponds to the dependence (\ref{temp}) obtained
previously using scaling arguments~\cite{TomThor}. 
 From Eqs.~(\ref{Difviatau},\ref{Tres},\ref{tfin}) follows the
time-dependence of the self-diffusion coefficient:
\begin{equation}
\frac{D(t)}{D_0}\!=\! \left(\!1+\!\frac{t}{\tau_0}
\right)^{\!\!-\frac{5}{6}}\!\!\!\!\! + a_3\,\delta 
\left(\!1\!+\frac{t}{\tau_0}\right)^{\!\!-1}\!\!\!\!\!+ a_4 
\delta^2 \,\left(\!1\!+\frac{t}{\tau_0}
\right)^{\!\!-\frac{7}{6}}\!\!\!\!\! + \cdots
\label{Dres}
\end{equation}
with pure numbers $a_3$ and $a_4$~\cite{constants} and with
\begin{equation}
D_0= \frac83 \pi^{1/2}\sigma^2 g_2(\sigma) n  T_0^{1/2} \, .
\end{equation}
Correspondingly, the mean-square displacement reads
asymptotically at $\tau_0 \ll t$:
\begin{equation}
\left\langle \left( \Delta r(t) \right)^2  \right\rangle  \sim 
t^{1/6}+a_3 \delta \,\log t \, .
\label{dRasym}
\end{equation}
This dependence holds true for time 
\begin{equation}
\tau_c(0)\,\delta^{-1} \ll t \ll \tau_c(0)\, \delta^{-11/5}\,,   
\end{equation}
where the
first inequality follows from the condition $\tau_0 \ll t$, while the
second one follows from the discussed condition $\tau_c(t) \ll
\tau_0$.  For the constant restitution coefficient one obtains
\begin{equation}
T(t)/T_0=\left[ 1+\gamma_0 t/\tau_c(0) \right]^{-2}\,,  
\end{equation}
where $\gamma_0 \equiv (1-\epsilon^2)/6$ \cite{GZ93,NEBO97}. Thus using 
Eqs.(\ref{tEt}) and  (\ref{Difviatau}) one obtains for the
mean-square displacement in this case
\begin{equation}
\label{drcons}
\left< \left( \Delta r(t) \right)^2
\right>  \sim \log t \, .
\end{equation}
As it follows from Eqs.(\ref{dRasym}) and (\ref{drcons}) the impact-velocity 
dependent restitution coefficient (\ref{epsilon}) leads to a significant 
change of the long-time behavior of the mean-square displacement of 
particles in the laboratory-time. Compared to its logarithmically 
weak dependence for the constant restitution coefficient, the 
impact-velocity dependence of the restitution coefficient (\ref{epsilon}) 
gives rise to a considerably faster increase of this quantity with time, 
according to a power law. 

\section{Results and discussion}

We studied the diffusion of particles in a homogeneously cooling granular 
gas. With the assumption of molecular chaos we calculated the velocity 
time-correlation function and self-diffusion coefficient. For 
impact-velocity dependent restitution coefficient we found a relation which 
expresses the diffusion coefficient in terms of material constants of 
particles and characteristics of the granular gas, such as temperature, 
density, etc.

In our calculations we used the restitution coefficient for viscoelastic collision of particles, which depends on the normal component of the impact velocity 
$|\vec{v}_{ij}\vec{e}|$ as a series $\sum |\vec{v}_{ij}\vec{e}|^{\gamma}$ 
with $\gamma=1/5$, $\gamma=2/5, \ldots$ [Eq.(\ref{epsilon})]. Our approach 
can be extended to arbitrary exponents $\gamma$, which let the expansion 
(\ref{epsilon}) converge. Using these exponents one may possibly describe
collisions with very large impact velocities (when plastic deformation
occurs) or very small velocities (when the surface effects are important), 
provided fragmentation/coagulation is ignored. 

For granular particles suffering viscoelastic collisions we found that the 
mean-square displacement grows with time as a power law $\sim t^{1/6}$, 
i.e. much faster than the logarithmic growth $\sim \log t$, observed in 
granular gases  with a constant restitution coefficient. It worth to note that 
qualitatively this power-law dependence (as well as the logarithmic one) 
simply follows from scaling arguments and the time-dependence of temperature. 
Indeed, the average velocity scales as $\bar{v} \sim T^{1/2}$, and therefore 
as $\sim t^{-1}$ for the constant restitution coefficient and as 
$\sim t^{-5/6}$ for the impact-velocity dependent coefficient. The diffusion 
coefficient in granular gas scales as $D \sim l^2/\tau_c$, where 
$l \sim \sigma^{-2} n^{-1}$ is the mean-free path, which does not change with 
time (in the regime preceding clustering), and $\tau_c \sim l/\bar{v}$ is the 
mean-collision time. Thus, $D \sim l \bar{v} \sim T^{1/2}$, and we obtain 
that the mean-square displacement, $\int^t D(t)dt$, scales as $\sim \log t$ 
in the former case and as $\sim t^{1/6}$ in the latter case.

What will be the impact of this apparently dramatic difference in the
time dependence of $\left< \left( \Delta r (t) \right)^2 \right> $ on
the properties of the granular gases? In the laboratory-time this
corresponds to the enhanced spreading (and therefore mixing) of
particles with the velocity-dependent restitution coefficient as
compared with the case of constant $\epsilon$. Since the temperature
decreases more slowly for the former case, as $\sim t^{-5/3}$,
as compared with $\sim t^{-2}$ for the latter one, retarded clustering may
be expected.

One might wish to define the average cumulative number of collisions per 
particle ${\cal N}(t)$ as a system inherent time-scale ($t$ is the 
laboratory-time) and compare dynamics of the systems in their 
inherent-time scale. ${\cal N}(t)$ is easily accessible in numerical simulations
and is a convenient quantity to analyze evolution of granular gases 
(e.g. \cite{EBEuro,McNamara96}). It may be found by integrating 
$d{\cal N}=\tau_c(t)^{-1}dt$ \cite{NEBO97}. For constant restitution 
coefficient one obtains ${\cal N}(t) \sim \log t$, while for the impact-velocity 
dependent ${\cal N}(t) \sim t^{1/6}$. Therefore, the mean-square
displacement behaves identically in both cases as

\begin{equation}
\label{drN}
\left< \left( \Delta r({\cal N}) \right)^2
\right>  \sim {\cal N} \, .
\end{equation}
If the number of collisions per particle ${\cal N}(t)$ was the only
quantity specifying the stage of the granular gas evolution, one would
conjecture that the dynamical behavior of a granular gas with a
constant $\epsilon$ and velocity-dependent $\epsilon$ is identical
provided ${\cal N}$-based time-scale is used.  According to our
understanding, however, this is not a adequate description of physical
reality. Indeed, as it was shown in Ref.\cite{EBEuro}, the value of
${\cal N}_c$, corresponding to a crossover from the linear regime of
evolution (which refers to the homogeneous cooling) to nonlinear
regime (when clustering starts) may differ by orders of magnitude,
depending on the restitution coefficient and on density of the granular
gas.  Therefore, to analyze the behavior of granular gas one can try
an alternative inherent time-scale, ${\cal T}^{-1} \equiv T(t)/T_0$
which is based on the gas temperature \cite{remcollapse}. Given two
systems of granular particles at the same density and the same initial
temperature $T_0$, consisting of particles colliding with constant and
velocity-dependent restitution coefficient, respectively, the time
${\cal T}$ allows to compare directly their evolution. A strong
argument to use a temperature-based time has been given by Goldhirsch
and Zanetti \cite{GZ93}: They found that there are two main
contributions to the rate of temperature decay.  The first one
refers to cooling due to inelastic collisions with a rate $\sim
T^{3/2}$ (for constant $\epsilon$) and corresponds to the homogeneous
cooling. The second one refers to viscous heating and behaves as $\sim
T^{1/2}$.  Initially the first is much larger than the second, but
when temperature decays, the second contribution takes over and the
shear wave adiabatically enslaves the temperature field \cite{GZ93}.
This corresponds to the nonlinear stage of evolution, when clustering
starts. Thus, the temperature may indicate the stage of evolution of a
granular gas. The recent numerical results of Ref.\cite{EBEuro}
strongly support our assumption: It was shown that while ${\cal N}_c$
differs by more than a factor three for two different systems (${\cal
  N}_c=70$ for a system with $\epsilon=0.9$ and packing fraction
$\phi=0.245$, and ${\cal N}_c=23$ for $\epsilon=0.6$ and $\phi=0.05$
\cite{EBEuro}) the values of ${\cal T}_c$, (defined, as $T({\cal
  N}_c)/T_0$) are very close (${\cal T}_c \approx 0.0031$ for the
first system and ${\cal T}_c \approx 0.0027$ for the second
\cite{EBEuro}).  These arguments show that one could consider ${\cal
  T}$ as a relevant time-scale to analyze the granular gas evolution.

With $T({\cal N})/T_0 \sim e^{-2 \gamma_0{\cal N}}$ for constant 
restitution coefficient and $T({\cal N})/T_0 \sim {\cal N}^{-10}$ for the 
impact-velocity dependent one, we obtain 
$\left< \left( \Delta r \right)^2\right>  \sim \log {\cal T}$ for the 
case of constant $\epsilon$ and
\begin{equation}
\label{drT}
\left< \left( \Delta r \right)^2 \right> \sim {\cal T}^{1/10}
\end{equation}
for the velocity-dependent restitution coefficient. Thus, for the 
temperature-based inherent time-scale we again obtain a power-law
dependence for the latter system and logarithmically-weak time-dependence
for the former one.

In conclusion, we found that that the impact-velocity dependence of the
restitution coefficient $\epsilon$ significantly influences the mean-square
displacement of the particles in granular gas in laboratory-time. As 
compared with the
logarithmically weak time-dependence found for a constant restitution
coefficient, the impact-velocity depending coefficient~(\ref{epsilon})
yields a power-law (\ref{dRasym}). It causes increasingly enhanced spreading 
of the particles through the system. This together with the fact that the 
temperature decreases more slowly for the velocity-dependent $\epsilon$ 
leads to the suggestion of retarded clustering in such systems. 

We also analyzed mean-square displacement of particles using different
inherent time-scales. We found, that while for the number-of-collision
based time-scale both systems behave identically, for the
temperature-based time-scale still power-law behavior is observed for
the case of velocity-dependent coefficient and logarithmically-weak
behavior for the case of constant $\epsilon$.

We thank M. H. Ernst and I. Goldhirsch for valuable discussions.

\end{multicols}

\begin{references}
\bibitem{remhydcon} The hydrodynamic contribution to the
  self-diffusion coefficient grows with time (as a time-integral of
  the time-correlation function). For granular materials it may be
  much more important than for fluids due to long-ranged spatial
  correlations in the velocity field \cite{NEBO97}.  
  This problem is beyond the scope of the present study.
\bibitem{pseudo} The term ``pseudo'' was initially referred to the
  dynamics of systems with singular hard-core potential
  \cite{Ernst:69,resibua}.
\bibitem{Chandler} For application to ``ordinary'' fluids, see
  \cite{Chandler1} and to granular systems
  \cite{NoijeErnst:97,HuthmannZippelius:97}.  A rigorous
  definition of ${\cal L}$ includes a pre-factor, preventing
  successive collisions of the same pair of particles
  \cite{Ernst:69,resibua} which, however, does not affect the present
  analysis.
\bibitem{expansion} Note that (\ref{tfin}) is an expansion in terms of
  $A \alpha^{2/5}T^{1/10}$. Thus, it restricts the temperature, $T$,
  to be small as compared with $A^{-10} \alpha^{-4}$.
\bibitem{remcollapse}
  Another problem of using the number of collision as an inherent
  time is connected with the phenomenon of inelastic collapse 
  \cite{McNamara} where an infinite number of collisions occurs in 
  finite time:~Suppose three particles of a (finite) system undergo 
  an inelastic collapse at time $t^*$, hence, ${\cal N}(t \to t^*) \to \infty$.
  This situation may occur in a homogeneously cooling system. If we chose 
  ${\cal N}$ to be the system inherent time, this time scale does not 
  allow to describe the evolution of the system for times larger than 
  $t^*$, whereas the temperature-based time-scale ${\cal T}$ is unaffected
  by the collapse.
\end{references}
\end{document}